\documentclass[sigconf, nonacm]{acmart}
\usepackage{caption}
\usepackage{subcaption}
\usepackage{float}

\AtBeginDocument{%
  \providecommand\BibTeX{{%
    \normalfont B\kern-0.5em{\scshape i\kern-0.25em b}\kern-0.8em\TeX}}}

\begin{document}

\title{GeoWINE: Geolocation based Wiki, Image, \\News and Event Retrieval}

\author{Golsa Tahmasebzadeh}
\affiliation{%
  \institution{ TIB -- Leibniz Information Centre for Science and Technology}
  \country{Germany}
}
\email{golsa.tahmasebzadeh@tib.eu}

\author{Endri Kacupaj}
\affiliation{%
  \institution{University of Bonn}
  \country{Germany}
}
\email{kacupaj@cs.uni-bonn.de}

\author{Eric M\"uller-Budack}
\affiliation{%
  \institution{ TIB -- Leibniz Information Centre for Science and Technology}
  \country{Germany}
}
\email{eric.mueller@tib.eu}

\author{Sherzod Hakimov}
\affiliation{%
  \institution{ TIB -- Leibniz Information Centre for Science and Technology}
  \country{Germany}
}
\email{sherzod.hakimov@tib.eu}

\author{Jens Lehmann}
\affiliation{%
  \institution{University of Bonn \\ Fraunhofer IAIS Dresden}
  \country{Germany}
}
\email{jens.lehmann@cs.uni-bonn.de}

\author{Ralph Ewerth}
\affiliation{%
  \institution{ TIB Hannover \& L3S Research Center, \\ Leibniz University Hannover}
  \country{Germany}
}
\email{ralph.ewerth@tib.eu}

\renewcommand{\shortauthors}{Tahmasebzadeh et al.}

\begin{abstract}
  In the context of social media, geolocation inference on news or events has become a very important task. In this paper, we present the GeoWINE (Geolocation-based Wiki-Image-News-Event retrieval) demonstrator, an effective modular system for multimodal retrieval which expects only a single image as input. The GeoWINE system consists of five modules in order to retrieve related information from various sources. The first module is a state-of-the-art model for geolocation estimation of images. The second module performs a geospatial-based query for entity retrieval using the Wikidata knowledge graph. The third module exploits four different image embedding representations, which are used to retrieve most similar entities compared to the input image. The embeddings are derived from the tasks of geolocation estimation, place recognition, ImageNet-based image classification, and their combination. The last two modules perform news and event retrieval from EventRegistry and the Open Event Knowledge Graph (OEKG). GeoWINE provides an intuitive interface for end-users and is insightful for experts for reconfiguration to individual setups. The GeoWINE achieves promising results in entity label prediction for images on Google Landmarks dataset. The demonstrator is publicly available at \url{http://cleopatra.ijs.si/geowine/}.
\end{abstract}


\maketitle

\section{Introduction} 
\label{sec:introduction}
\paragraph{\textbf{Motivation.}}

With the rapid growth of technology, billions of images are appearing on the Internet every day. While surfing the Web one may encounter an image (e.g., depicting a place) and would like to know where it belongs to and what are similar places. Nonetheless, it would be hard to find keywords to query a search engine. With this in mind, reverse-image search is a unique way to browse the Web. Some of existing reverse-image search implementations are Google Images, eBay, Bing, Pinterest, and Alibaba, most of which are mainly based on extracting visual features from the query image or meta information to perform the retrieval task~\cite{al2020awjedni}. In this regard, content-based location estimation of images would enhance the retrieval task. There is an increasing research interest in exploiting geographical content of images for various tasks like image classification~\cite{feifeiliclassify}, information retrieval~\cite{mlm}, and image verification~\cite{QATM}. On the other hand, information retrieval, text and Web mining, as well as knowledge graphs are increasingly intertwined with the purpose of yielding opportunities for enhanced exploration facilities. 

 In this paper, we propose GeoWINE (Geolocation Wiki-Image-News-Events) to close the gaps between geolocation estimation, information representation in knowledge graphs, and information retrieval. We utilize geolocation estimation as a core task, but also go beyond by including contextual information in order to enrich the information retrieval task. In this regard, the proposed system is an extension of geolocation estimation to highlight its potential by linking the input image to other knowledge sources such as Wikidata~\cite{wikidata} and Open Event Knowledge Graph (OEKG)~\cite{oekg}. To the best of our knowledge, this is the first attempt based on image-based geolocation 
 estimation for event and news retrieval in the context of more downstream tasks such as image verification, recommending similar places, or fact-checking.

\begin{figure*}[!t]
  \centering
  \includegraphics[width=\linewidth]{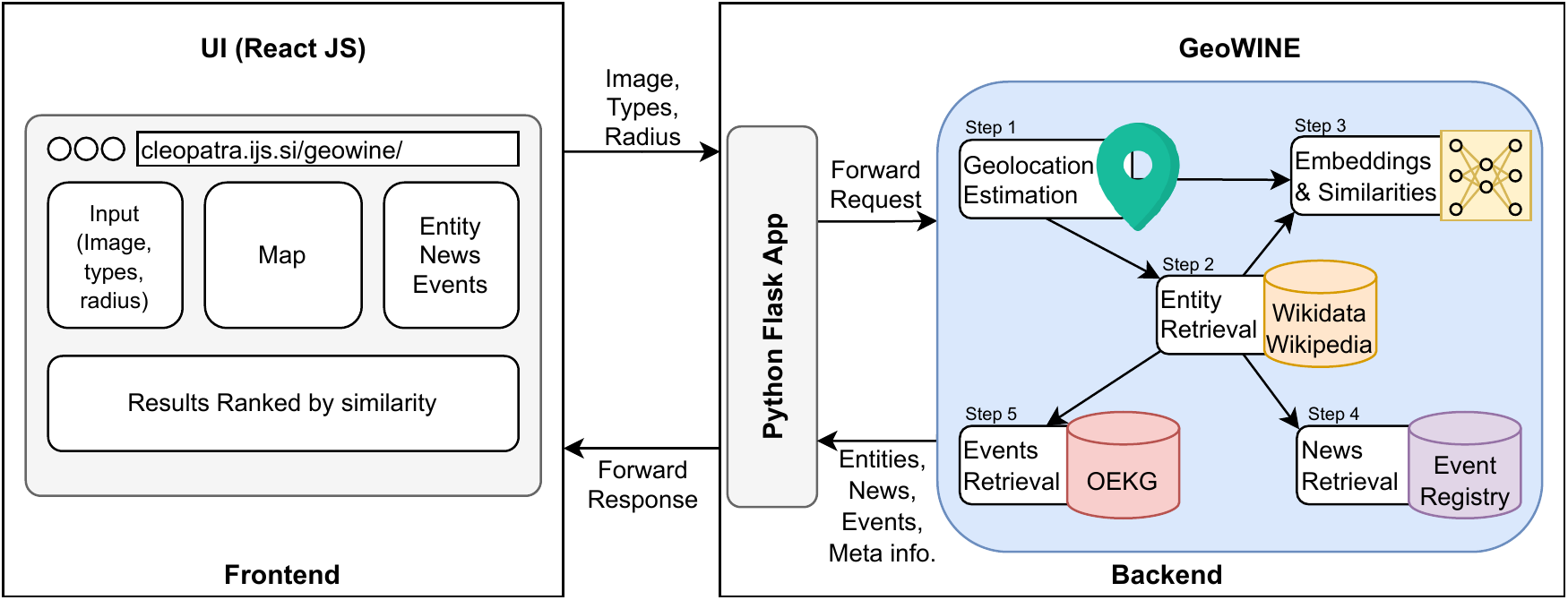}
  \caption{Overview of the GeoWINE architecture.}
  \label{fig:geowine}
  \vspace{-1em}
\end{figure*}

\paragraph{\textbf{Approach and Contribution.}}
The proposed system is a geolocation based multimodal retrieval system that comprises five modules. Given an image as input, GeoWINE applies a state-of-the-art geolocation estimation model as a starting point to retrieve data from Wikidata~\cite{wikidata}, EventRegistry\footnote{\url{http://eventregistry.org/}}, and the OEKG~\cite{oekg}. The geolocation estimation model is responsible for the prediction of the input image coordinates. The second module performs a geospatial query on Wikidata to retrieve all entities of specific types located no farther than a specified radius from the predicted coordinates. Here, the entity types and the radius are given as input to our system. The third module leverages three different image embedding representations that are derived from the tasks of geolocation estimation, place recognition, and an ImageNet model for image classification. These embeddings are used to find and rank the most similar entities compared to the input image. Finally, the last two modules retrieve similar news and events from EventRegistry and OEKG. 

We evaluate GeoWINE on Google Landmarks dataset~\cite{googlelandmarkdataset}, where it achieves promising performance in predicting entity labels of query images.
GeoWINE enables users to retrieve entities, news, and events related to an image, through a clean and intuitive User Interface (UI), with interactive response times. To the best of our knowledge, this is the first public and open-source demo for geolocation-based multimodal retrieval through various sources. To facilitate reproducibility and reuse, all material is publicly available\footnote{\url{https://github.com/cleopatra-itn/GeoWINE}}.

\paragraph{\textbf{Use cases.}}
The proposed system provides various search scenarios each of which could be addressed as a use case. One of the use cases is to assist journalists or social scientists in fact-checking or fake news detection~\cite{fakenews}. Given a query image as an input, the system could be used to contextualize the image by retrieving relevant documents. First, the geolocation is predicted and then the predicted coordinates are used to query Wikidata~\cite{wikidata} to retrieve entities within the selected radius and entity types by the user. In this way, the user can select an entity on the map to get back the relevant news and events in order to check whether the input image is relevant to the retrieved documents or their corresponding images. Another useful application is the verification of images based on the retrieved entities. For instance, given an image of a scenery, the system recommends a ranked list of visually similar places within a selected radius in order to verify the label of the input image. From another perspective the system could also be useful in meta information verification of news. For instance to see if the location claimed of a news matches the location of its image.

\section{Method} 
\label{sec:method}
In this section, we describe the five modules of the GeoWINE system. The basic functionality is shown in Figure~\ref{fig:geowine}.

\subsection{Geolocation Estimation Module} \label{sec:geoestim}
The goal of geolocation estimation is to predict latitude and longitude coordinates where photo was captured, using only the visual content without any other metadata. We apply the model\footnote{\url{https://github.com/TIBHannover/GeoEstimation/tree/pytorch}}~\cite{hierarchcal} based on ResNet101~\cite{resnet_v2} pretrained on a subset of the Yahoo Flickr Creative Commons 100 Million dataset (YFCC100M)~\cite{yfcc} with around five million geo-tagged images. In this model, geolocation estimation is treated as a classification problem where the earth is subdivided into geographical cells. This approach exploits the hierarchical representation of multiple partitionings and additionally takes the photo’s scene category into account, i.e., whether it depicts indoor, natural, or urban setting.

\subsection{Geospatial-based Entity Retrieval Module}
The primary knowledge source for our system is the Wikidata knowledge graph~\cite{wikidata} which is a free and open knowledge base that can be read and edited by both humans and machines. Since Wikidata is a rich source for data with semantic relations between entities and gives continuous access using public SPARQL (Simple Protocol and RDF Query Language) endpoint, it meets our requirements for entity retrieval.

\subsubsection{Determining Common Entity Types}\label{sec:entity_group}
In order to determine entity types for the retrieval task, we use an index set from the Google Landmarks dataset~\cite{googlelandmarkdataset}. Each image in the index set is linked with the corresponding Wikimedia Commons category, which is also linked with a Wikidata entity. We query the types of Wikidata entities using the \textit{instance of} (P31) of each image and count their occurrences. The most frequent 35 entity types are selected and clustered in 12 groups where each group contains similar entity types. For instance, \textit{Religious building} contains entity types such as \textit{church building}, \textit{cathedral}, \textit{Buddhist temple}, \textit{shrine}, etc. The full list of entity groups is given in Table~\ref{tab:evaluation}.


\subsubsection{Querying Entities from Wikidata} \label{sec:query_entities}
The geolocation estimation module (Section~\ref{sec:geoestim}) outputs the geocoordinates for an input image. In this step, we query Wikidata to retrieve entities of the specified type and located in the selected radius (in kilometers) of the predicted geocoordinates. The user specifies the radius and selects the entity types of interest from the available 12 groups.

\subsubsection{Retrieval of Metadata and Images}Eventually, corresponding text descriptions and images are downloaded for the retrieved entities. We extract Wikimedia Commons links of the images and the Wikipedia link for each queried entity using MediaWiki and Wikidata query APIs (Application Programming Interface)\footnote{\url{https://www.mediawiki.org/wiki/API:Main_page}}.

\subsection{Visual Search Embeddings Module} \label{sec:embeddings}

To generate the visual representations, state-of-the-art approaches in computer vision are applied. The \textit{Geolocation} embedding is taken from the same model used for geolocation estimation~\cite{hierarchcal} (Section~\ref{sec:geoestim}). The \textit{Place} embedding is another representation based on a ResNet model~\cite{resnet_v2} to recognize 365 distinct places like \textit{beach} or \textit{street}. Another embedding applied here is \textit{Object} embedding based on a ResNet model pre-trained on the ImageNet dataset for image classification~\cite{resnet}. Each of the embeddings is encoded in a vector of size 2048. Eventually, the \textit{Combined} is resulted from concatenation of all three to get a more general visual representation. These representations are employed to rank the retrieved entities (Section~\ref{sec:query_entities}) based on similarity to the input query image. The similarity between the entity image and the query image is computed using the cosine similarity of their corresponding image embeddings.

\subsection{News and Event Retrieval Modules}
Given the query image, the user can select the corresponding news articles or events for the retrieved entities within the defined radius. 

We use EventRegistry to retrieve news for each entity, which is a repository of news in different domains, locations and languages. To get relevant news for each selected entity, its Wikidata label is used as a keyword to query the relevant news.

Regarding the events, we use OEKG~\cite{oekg}, which is a multilingual, event-centric, temporal knowledge graph composed of different datasets linked to Event-centric Knowledge Graph (EventKG)~\cite{eventkg}. These datasets come from multiple application domains such as question answering, entity recommendation, and named entity recognition. We use the OEKG endpoint\footnote{\url{http://oekg.l3s.uni-hannover.de/sparql}} to query the relevant events from the knowledge graph for the selected Wikidata entity.

\section{\textbf{System Overview}}
\label{sec:system_overview}

An overview of the proposed system is shown in Figure~\ref{fig:geowine}. The demonstrator consists of a frontend and a Python Flask server as the backend.

\subsection{Frontend}
The frontend was built using React JS (JavaScript), an open-source JavaScript library for user interface components. The page comprises four main panels: 1) the panel to provide the query image (either by selecting an existing one or uploading new images) alongside the entity type and radius, 2) the map that displays the \textit{entity results}, 3) the panel that provides entity details, news and events, and 4) the ordered entities based on the similarity to the input image. Once the user presses the search button, the provided image, the selected entity type, and the defined radius are sent to the backend.

\subsection{Backend}
In the backend, the request is sent to a Python Flask application in JSON (JavaScript Object Notation) format. Then, the application forwards the request to an initialized instance of GeoWINE, which computes the results using the pipeline components specified in Section~\ref{sec:method}. Finally, the Flask application sends the computed results back to the frontend in JSON format.

\begin{figure}[!t]
  \centering
  \includegraphics[width=\columnwidth]{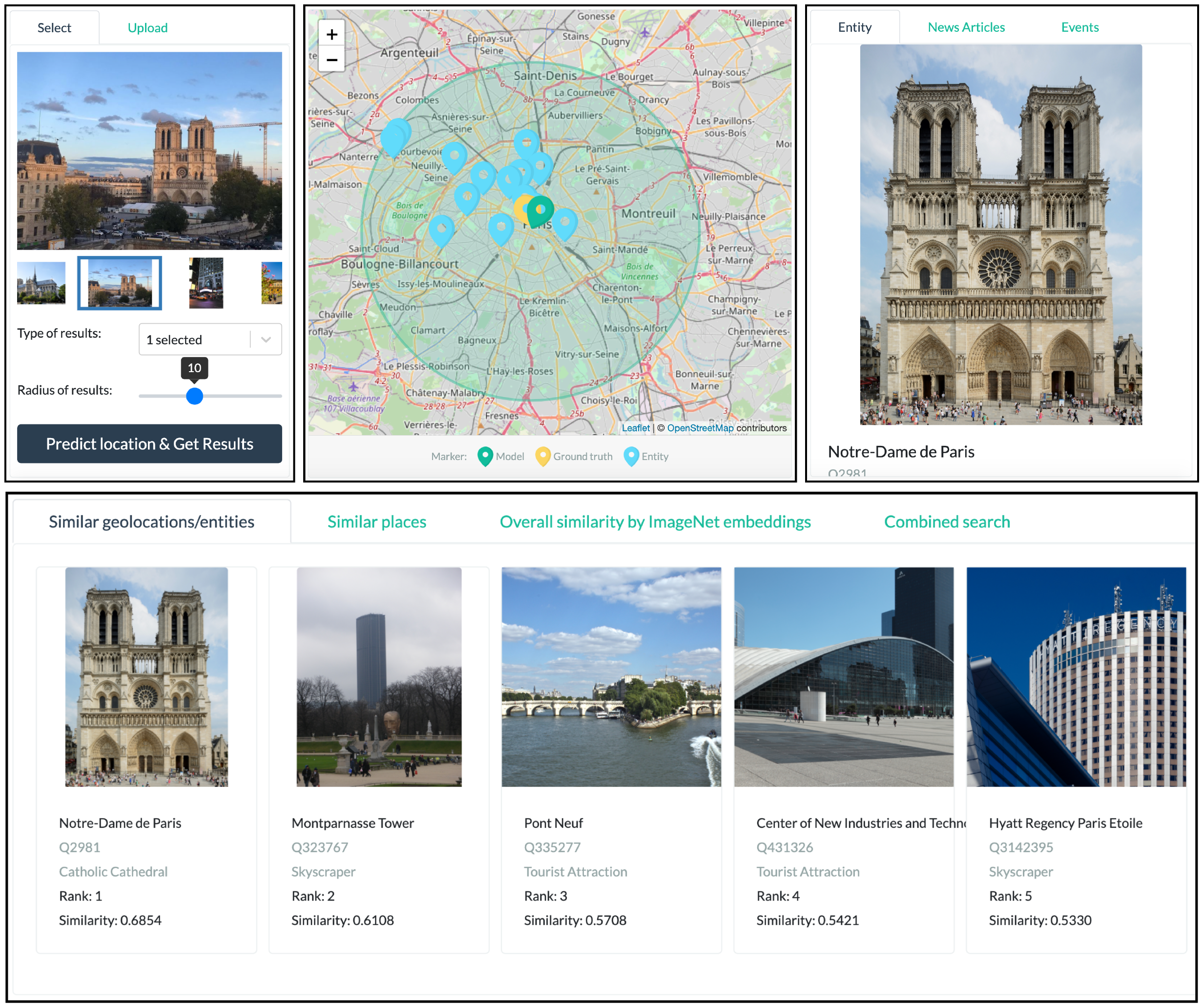}
  \caption{Snapshots of the demonstration on a sample image.}
  \label{fig:demo}
\end{figure}

\begin{table*}
\caption{The accuracy scores at $k$ ranked results (in [\%]) and the number of instances for different input entity groups: Bridge (Br), Historic Site (Hi), Square (Sq), Castle (Ca), Monument (Mo), Museum (Mu), Building (Bu), Religious Building (Rb), Tower (To), Tourist Attraction (Ta), Waterfall (Wa), Skyscraper (Sk).}
\label{tab:evaluation}
\begin{center}
\begin{tabular}{lccccccccccccccl}
\toprule
\small
\centering
\textbf{in [\%]} & \textbf{Br} & \textbf{Hi} & \textbf{Sq} & \textbf{Ca} & \textbf{Mo} &\textbf{Mu} & \textbf{Bu} & \textbf{Rb} & \textbf{To} & \textbf{Ta} & \textbf{Wa} & \textbf{Sk} & \textbf{overall}\\
\hline
\textbf{Top 10} & \textbf{90} & 79 &  73 & 86 & 72 & 67 &  86 & 50 &  77 & 81  & \textbf{90} & 65 & 84 \\
\textbf{Top 5} & 74 & 70 &  64 & 77 & 68 & 60 & \textbf{86} &  50 & 77  & 81 & 85 & 55 & 76\\
\textbf{Top 1} & 53 & 52 & 27 & 50 & 56 & 40 & 57 & 13 & 46 & 68 & \textbf{75} & 35 & 52 \\
\hline
\textbf{\# instances per entity group}& 19 & 33 &  33 & 22 & 25 & 30 & 14 & 8 &  13 & 31 & 20 & 40 & 275 \\

\bottomrule
\end{tabular}
\end{center}
\vspace{-1em}
\end{table*}

\section{\textbf{Demo Walkthrough}}
\label{sec:walkthrough}
In this section, we provide an example query in Figure~\ref{fig:demo} to guide the reader and user through the demo\footnote{\url{http://cleopatra.ijs.si/geowine/}}. 

Users can upload an image or select from the provided list, specify a radius in kilometers, and select entity types (explained in Section~\ref{sec:entity_group}) from the drop-down list and then get the results as illustrated in the top left box. Given these inputs, the geocoordinates are predicted and pinpointed on the map (middle box). In addition to the predicted location (the green marker), the corresponding Wikidata entities are represented on the map within the selected radius (blue markers). For the pre-selected images the ground truth geocoordinates are available so that the user can compare the performance of the system. In this example, the location is predicted correctly, and the label is \textit{Notre-Dame Paris} (top right box).

In addition to the aforementioned outputs, a list of images is presented at the bottom. These images correspond to the retrieved entities on the map and they are ranked based on similarity to the input image. There are four different options here: similar geolocations/entities, similar places, overall similarity by ImageNet embeddings, and combined search. These four different ranked lists correspond to the four different visual representations: \textit{Geolocation}, \textit{Place}, \textit{Object} and \textit{Combined} respectively (Section~\ref{sec:embeddings}). As shown in Figure~\ref{fig:demo} bottom box, the top-1 similar geolocation is \textit{Notre-Dame Paris}. Eventually, the user can select one of the retrieved entities on the map and see the relevant news and events (top right box). 

\section{Evaluation}
\label{sec:evaluation}
In this section, we present an evaluation of the presented demonstrator on a subset of the Google Landmarks dataset. 

As mentioned in Section~\ref{sec:entity_group}, we identified the most frequent entity types and formed 12 distinct groups by clustering the types of similar imagery. Then, we selected one type from each group and downloaded the images, which yielded $13280$ images in total. The goal is to see if the system is able to retrieve the corresponding Wikidata~\cite{wikidata} entity when the Wikimedia Commons image is given as the query. Initially, we have run the geolocation estimation model on the query images and discarded query images whose estimated geocoordinate errors are above city level (25 kilometres) based on Great Circle Distance (GCD) in comparison with the ground truth geocoordinates. From the remaining $1285$ images, we extract geolocation embeddings and rank the retrieved entities based on the similarity to the input image. In order to make the evaluation fair, we discarded all the queries where the image was the same as the Wikidata image by comparing the hash codes of the query and the retrieved images. 

For the remaining 275 query images, the top-$k$ accuracy is given in Table~\ref{tab:evaluation} for different input entity groups. As the Table shows, 84\%, 76\%, and 52\% of the test queries are ranked among the top 10, top 5, and top 1, respectively. Among the top-10 results, \textit{Waterfall} and \textit{Bridge}, among the top-5 \textit{Building} and among top-1 \textit{Waterfall} entity groups have the highest accuracy. In all the top-k results, \textit{Religious building} achieves the lowest accuracy. Overall, the experimental results show the feasibility and usability of the whole system for different types of landmarks.

\section{Conclusions}
In this paper, we have introduced the GeoWINE demonstrator, a geolocation-based multimodal retrieval system aimed at supporting different kinds of applications such as fact-checking, fake news detection, or image verification. The system consists of five modules: geolocation estimation, geospatial-based entity retrieval, visual search embeddings, news and event retrieval. The user can provide an image, radius, and types of results (e.g. tourist attraction) as input and receive the image geocoordinates on the map, surrounding entities from Wikidata, and visually similar images from the retrieved entities. Furthermore, the user can find some relevant news and events based on a selected entity on the map. In order to evaluate the performance of the system, we have used a small subset of the Google Landmark dataset for labeling images. The experiments have shown promising results affirming the potential of the overall system for the applications mentioned above.

\begin{acks}
This work was supported by the European Union H2020 founded project CLEOPATRA (ITN, GA. 812997).
\end{acks}

\bibliographystyle{acmbib}
\bibliography{main}


\begin{thebibliography}{13}


\ifx \showCODEN    \undefined \def \showCODEN     #1{\unskip}     \fi
\ifx \showDOI      \undefined \def \showDOI       #1{#1}\fi
\ifx \showISBNx    \undefined \def \showISBNx     #1{\unskip}     \fi
\ifx \showISBNxiii \undefined \def \showISBNxiii  #1{\unskip}     \fi
\ifx \showISSN     \undefined \def \showISSN      #1{\unskip}     \fi
\ifx \showLCCN     \undefined \def \showLCCN      #1{\unskip}     \fi
\ifx \shownote     \undefined \def \shownote      #1{#1}          \fi
\ifx \showarticletitle \undefined \def \showarticletitle #1{#1}   \fi
\ifx \showURL      \undefined \def \showURL       {\relax}        \fi
\providecommand\bibfield[2]{#2}
\providecommand\bibinfo[2]{#2}
\providecommand\natexlab[1]{#1}
\providecommand\showeprint[2][]{arXiv:#2}

\bibitem[\protect\citeauthoryear{Al-Lohibi, Alkhamisi, Assagran, Aljohani, and
  Aljahdali}{Al-Lohibi et~al\mbox{.}}{2020}]%
        {al2020awjedni}
\bibfield{author}{\bibinfo{person}{Hanaa Al-Lohibi}, \bibinfo{person}{Tahani
  Alkhamisi}, \bibinfo{person}{Maha Assagran}, \bibinfo{person}{Amal Aljohani},
  {and} \bibinfo{person}{Asia~Othaman Aljahdali}.}
  \bibinfo{year}{2020}\natexlab{}.
\newblock \showarticletitle{Awjedni: A Reverse-Image-Search Application}.
\newblock \bibinfo{journal}{\emph{ADCAIJ}} (\bibinfo{year}{2020}).
\newblock


\bibitem[\protect\citeauthoryear{Armitage, Kacupaj, Tahmasebzadeh, Swati,
  Maleshkova, Ewerth, and Lehmann}{Armitage et~al\mbox{.}}{2020}]%
        {mlm}
\bibfield{author}{\bibinfo{person}{Jason Armitage}, \bibinfo{person}{Endri
  Kacupaj}, \bibinfo{person}{Golsa Tahmasebzadeh}, \bibinfo{person}{Swati},
  \bibinfo{person}{Maria Maleshkova}, \bibinfo{person}{Ralph Ewerth}, {and}
  \bibinfo{person}{Jens Lehmann}.} \bibinfo{year}{2020}\natexlab{}.
\newblock \showarticletitle{{MLM:} {A} Benchmark Dataset for Multitask Learning
  with Multiple Languages and Modalities}. In \bibinfo{booktitle}{\emph{{CIKM}
  2020}}. \bibinfo{publisher}{{ACM}}.
\newblock


\bibitem[\protect\citeauthoryear{Cheng, Wu, AbdAlmageed, and Natarajan}{Cheng
  et~al\mbox{.}}{2019}]%
        {QATM}
\bibfield{author}{\bibinfo{person}{Jiaxin Cheng}, \bibinfo{person}{Yue Wu},
  \bibinfo{person}{Wael AbdAlmageed}, {and} \bibinfo{person}{Premkumar
  Natarajan}.} \bibinfo{year}{2019}\natexlab{}.
\newblock \showarticletitle{{QATM:} Quality-Aware Template Matching for Deep
  Learning}. In \bibinfo{booktitle}{\emph{{IEEE} Conference on Computer Vision
  and Pattern Recognition, {CVPR} 2019}}. \bibinfo{publisher}{Computer Vision
  Foundation / {IEEE}}.
\newblock


\bibitem[\protect\citeauthoryear{Gottschalk and Demidova}{Gottschalk and
  Demidova}{2019}]%
        {eventkg}
\bibfield{author}{\bibinfo{person}{Simon Gottschalk} {and}
  \bibinfo{person}{Elena Demidova}.} \bibinfo{year}{2019}\natexlab{}.
\newblock \showarticletitle{EventKG - the hub of event knowledge on the web -
  and biographical timeline generation}.
\newblock \bibinfo{journal}{\emph{Semantic Web}} (\bibinfo{year}{2019}).
\newblock


\bibitem[\protect\citeauthoryear{Gottschalk, Kacupaj, Abdollahi, Alves, Amaral,
  Koutsiana, Kuculo, Major, Mello, Cheema, Sittar, Swati, Tahmasebzadeh, and
  Thakkar}{Gottschalk et~al\mbox{.}}{2021}]%
        {oekg}
\bibfield{author}{\bibinfo{person}{Simon Gottschalk}, \bibinfo{person}{Endri
  Kacupaj}, \bibinfo{person}{Sara Abdollahi}, \bibinfo{person}{Diego Alves},
  \bibinfo{person}{Gabriel Amaral}, \bibinfo{person}{Elisavet Koutsiana},
  \bibinfo{person}{Tin Kuculo}, \bibinfo{person}{Daniela Major},
  \bibinfo{person}{Caio Mello}, \bibinfo{person}{Gullal~Singh Cheema},
  \bibinfo{person}{Abdul Sittar}, \bibinfo{person}{Swati},
  \bibinfo{person}{Golsa Tahmasebzadeh}, {and} \bibinfo{person}{Gaurish
  Thakkar}.} \bibinfo{year}{2021}\natexlab{}.
\newblock \showarticletitle{{OEKG: The Open Event Knowledge Graph}}. In
  \bibinfo{booktitle}{\emph{CLEOPATRA Workshop @ TheWebConf}}.
\newblock


\bibitem[\protect\citeauthoryear{He, Zhang, Ren, and Sun}{He
  et~al\mbox{.}}{2016a}]%
        {resnet}
\bibfield{author}{\bibinfo{person}{Kaiming He}, \bibinfo{person}{Xiangyu
  Zhang}, \bibinfo{person}{Shaoqing Ren}, {and} \bibinfo{person}{Jian Sun}.}
  \bibinfo{year}{2016}\natexlab{a}.
\newblock \showarticletitle{Deep Residual Learning for Image Recognition}. In
  \bibinfo{booktitle}{\emph{{CVPR} 2016}}. \bibinfo{publisher}{{IEEE} Computer
  Society}.
\newblock


\bibitem[\protect\citeauthoryear{He, Zhang, Ren, and Sun}{He
  et~al\mbox{.}}{2016b}]%
        {resnet_v2}
\bibfield{author}{\bibinfo{person}{Kaiming He}, \bibinfo{person}{Xiangyu
  Zhang}, \bibinfo{person}{Shaoqing Ren}, {and} \bibinfo{person}{Jian Sun}.}
  \bibinfo{year}{2016}\natexlab{b}.
\newblock \showarticletitle{Identity Mappings in Deep Residual Networks}. In
  \bibinfo{booktitle}{\emph{{ECCV} 2016}} \emph{(\bibinfo{series}{Lecture Notes
  in Computer Science})}. \bibinfo{publisher}{Springer}.
\newblock


\bibitem[\protect\citeauthoryear{M{\"{u}}ller{-}Budack, Pustu{-}Iren, and
  Ewerth}{M{\"{u}}ller{-}Budack et~al\mbox{.}}{2018}]%
        {hierarchcal}
\bibfield{author}{\bibinfo{person}{Eric M{\"{u}}ller{-}Budack},
  \bibinfo{person}{Kader Pustu{-}Iren}, {and} \bibinfo{person}{Ralph Ewerth}.}
  \bibinfo{year}{2018}\natexlab{}.
\newblock \showarticletitle{Geolocation Estimation of Photos Using a
  Hierarchical Model and Scene Classification}. In
  \bibinfo{booktitle}{\emph{{ECCV} 2018}} \emph{(\bibinfo{series}{Lecture Notes
  in Computer Science}, Vol.~\bibinfo{volume}{11216})}.
  \bibinfo{publisher}{Springer}, \bibinfo{pages}{575--592}.
\newblock


\bibitem[\protect\citeauthoryear{M{\"{u}}ller{-}Budack, Theiner, Diering,
  Idahl, and Ewerth}{M{\"{u}}ller{-}Budack et~al\mbox{.}}{2020}]%
        {fakenews}
\bibfield{author}{\bibinfo{person}{Eric M{\"{u}}ller{-}Budack},
  \bibinfo{person}{Jonas Theiner}, \bibinfo{person}{Sebastian Diering},
  \bibinfo{person}{Maximilian Idahl}, {and} \bibinfo{person}{Ralph Ewerth}.}
  \bibinfo{year}{2020}\natexlab{}.
\newblock \showarticletitle{Multimodal Analytics for Real-world News using
  Measures of Cross-modal Entity Consistency}.
\newblock \bibinfo{journal}{\emph{CoRR}} (\bibinfo{year}{2020}).
\newblock


\bibitem[\protect\citeauthoryear{Tang, Paluri, Li, Fergus, and Bourdev}{Tang
  et~al\mbox{.}}{2015}]%
        {feifeiliclassify}
\bibfield{author}{\bibinfo{person}{Kevin~D. Tang}, \bibinfo{person}{Manohar
  Paluri}, \bibinfo{person}{Fei{-}Fei Li}, \bibinfo{person}{Robert Fergus},
  {and} \bibinfo{person}{Lubomir~D. Bourdev}.} \bibinfo{year}{2015}\natexlab{}.
\newblock \showarticletitle{Improving Image Classification with Location
  Context}. In \bibinfo{booktitle}{\emph{{ICCV} 2015}}.
  \bibinfo{publisher}{{IEEE} Computer Society}.
\newblock


\bibitem[\protect\citeauthoryear{Thomee, Shamma, Friedland, Elizalde, Ni,
  Poland, Borth, and Li}{Thomee et~al\mbox{.}}{2016}]%
        {yfcc}
\bibfield{author}{\bibinfo{person}{Bart Thomee}, \bibinfo{person}{David~A.
  Shamma}, \bibinfo{person}{Gerald Friedland}, \bibinfo{person}{Benjamin
  Elizalde}, \bibinfo{person}{Karl Ni}, \bibinfo{person}{Douglas Poland},
  \bibinfo{person}{Damian Borth}, {and} \bibinfo{person}{Li{-}Jia Li}.}
  \bibinfo{year}{2016}\natexlab{}.
\newblock \showarticletitle{{YFCC100M:} the new data in multimedia research}.
\newblock \bibinfo{journal}{\emph{Commun. {ACM}}} (\bibinfo{year}{2016}).
\newblock


\bibitem[\protect\citeauthoryear{Vrandecic and Kr{\"{o}}tzsch}{Vrandecic and
  Kr{\"{o}}tzsch}{2014}]%
        {wikidata}
\bibfield{author}{\bibinfo{person}{Denny Vrandecic} {and}
  \bibinfo{person}{Markus Kr{\"{o}}tzsch}.} \bibinfo{year}{2014}\natexlab{}.
\newblock \showarticletitle{Wikidata: a free collaborative knowledgebase}.
\newblock \bibinfo{journal}{\emph{Commun. {ACM}}} (\bibinfo{year}{2014}).
\newblock


\bibitem[\protect\citeauthoryear{Weyand, Araujo, Cao, and Sim}{Weyand
  et~al\mbox{.}}{2020}]%
        {googlelandmarkdataset}
\bibfield{author}{\bibinfo{person}{Tobias Weyand}, \bibinfo{person}{Andr{\'{e}}
  Araujo}, \bibinfo{person}{Bingyi Cao}, {and} \bibinfo{person}{Jack Sim}.}
  \bibinfo{year}{2020}\natexlab{}.
\newblock \showarticletitle{Google Landmarks Dataset v2 - {A} Large-Scale
  Benchmark for Instance-Level Recognition and Retrieval}. In
  \bibinfo{booktitle}{\emph{{CVPR} 2020}}. \bibinfo{publisher}{{IEEE}}.
\newblock


\end{thebibliography}

\end{document}